\documentclass[a4paper,11pt]{article}
\usepackage{graphicx}
\usepackage{subcaption}
\usepackage{pos}
\usepackage{lipsum} 

\usepackage{lineno}

\title{Machine Learning Tools for the IceCube-Gen2 Optical Array}

\ShortTitle{Machine Learning Tools for the IceCube-Gen2 Optical Array}

\author{The IceCube-Gen2 Collaboration \\{\normalsize \normalfont(a complete list of authors can be found at the end of the proceedings)}\\}

\emailAdd{fvaracar@uni-muenster.de}
\emailAdd{jonas.selter@uni-muenster.de}

\abstract{

Neural networks (NNs) have a great potential for future neutrino telescopes such as IceCube-Gen2, the planned high-energy extension of the IceCube observatory. IceCube-Gen2 will feature new optical sensors with multiple photomultiplier tubes (PMTs) designed to provide omnidirectional sensitivity. 
Neural networks excel at handling high-dimensional problems and can naturally incorporate the increased complexity of these new sensors. Additionally, their fast inference time makes them promising candidates for handling the high event rates expected from IceCube-Gen2.

This contribution presents potential applications of neural networks in the IceCube-Gen2 in-ice optical array. First, we introduce a method to simulate the IceCube-Gen2 optical modules’ photon acceptance using a NN that leverages the modules’ inherent symmetries. Secondly, we present the status of neutrino NN–based reconstruction efforts, including the adaptation of a novel IceCube technique that combines normalizing flows with transformer NNs. Finally, we describe current progress in noise cleaning applications based on node classification with graph neural networks (GNNs), a method that has already shown promising results for the forthcoming low-energy extension, IceCube-Upgrade.

\vspace{4mm}

{\bfseries Corresponding authors:}
Francisco Javier Vara Carbonell$^{1*}$, 
Jonas Selter$^{1}$,\\
{$^{1}$ \itshape Institut für Kernphysik, Universität Münster}\\[4mm]
$^*$ Presenter
}

\ConferenceLogo{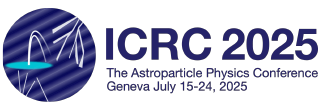}

\FullConference{39th International Cosmic Ray Conference (ICRC2025)\\
 15–24 July 2025\\
Geneva, Switzerland\\}

\begin{document}

\maketitle

\section{Introduction}\label{sec1}
IceCube-Gen2 \cite{ICRCGen2} will feature an optical array eight times larger than the current IceCube.\ For this new array, new digital optical modules (DOMs) are being developed \cite{ICRCLOM}.\ Although the final module design has not been decided yet, the current candidates share the use of several photomultiplier tubes (PMTs) embedded within a glass vessel, whose photocathodes are surrounded by passive optical components called gel pads to increase photon detection via total internal reflection.\ The new modules will provide better sensitivity than the current IceCube DOMs, with up to four times more effective area on average.\ However, new algorithms need to be developed for tasks such as event reconstruction or optical-module simulation in order to incorporate the increased complexity of these modules.\ In this context, deep learning techniques are promising candidates, as they excel at high-dimensional problems.

In the following, we present three potential neural-network applications for the IceCube-Gen2 optical array, using the 16-PMT prototype DOM as our baseline \cite{ICRCLOM}.\ This module houses sixteen 4-inch-diameter PMTs arranged symmetrically in upper and lower hemispheres, with each hemisphere further divided into four polar and four equatorial PMTs.

\section{Optical Module Simulation Using Neural Networks}

The IceCube simulation chain includes several consecutive steps, such as neutrino interaction, lepton propagation, and photon propagation through the ice.\ At the end of the last step, if a photon reaches the vicinity of an optical module, we must determine whether it will be detected by any of the PMTs and, if so, convert it into a photoelectron that continues through the rest of the simulation chain.\ In this work, we refer to this assignment of photons to specific PMTs as optical‐module simulation.

\subsection{Analytical Approximation}
At the moment, IceCube-Gen2 DOM simulation is implemented in the PPC framework \cite{PPC} using a plane wave approximation, which considers only the direction $\vec{n}$ of the photon.\ The angular response of each PMT within the DOM is parametrized with the same function $f(\beta, (\vec{m}\cdot\vec{n}))$ of the direction the PMT points in, $\vec{m}$, and a shape parameter $\beta$.\ In case of the IceCube-Gen2 DOM, $\beta$ and the elevation angles were obtained from fits to Geant4 simulations at a photon wavelength of 400\;nm.

While this approach preserves the expected DOM-level average photon acceptance, it cannot capture the full complexity of the new DOM. For instance, it enforces axial symmetry about each PMT axis, ignoring asymmetries from gel-pad geometry and PMT misalignment.\ Additionally, it assumes a wavelength-independent relative response, despite the wavelength-dependent optical properties of the glass and, more importantly, the gel.\ Finally, it treats polar and equatorial PMTs identically, even though they have different responses, mainly due to the different shapes of the gel pads.

Note that in the future this model might be improved by a description with two terms per PMT type (each with its own $\beta$ and elevation angle) to help address the first and last limitations.

\subsection{Neural Network Approach}

Given the limitations of the current analytical approximation, we pursued a neural network (NN) approach that can naturally accommodate the two PMT locations (polar and equatorial), the wavelength dependence of the angular acceptance, and its asymmetry.

In this approach, photons are propagated until they intersect a predefined geometrical surface around the optical module---for example, a sphere enclosing the IceCube-Gen2 DOM.\ The neural network then receives the photon’s landing position on this surface, its wavelength, and its direction as inputs, and outputs the detection probability for each PMT.

The network is trained on Geant4 simulations produced with the OMSim framework \cite{OMSim_1_2_0}.\ Roughly 20 billion individual photons were generated isotropically from a predefined spherical surface, with a flat wavelength distribution from 270\;nm to 700\;nm.\ OMSim assigns each photon a detection probability derived from the DOM’s measured optical properties and from position-dependent detection efficiency calibrated to PMT lab measurements and the Geant4 simulation.

The model was created using the OMNNSim framework \cite{VaraCarbonell_OMNNSim} and contains two branches:
\begin{itemize}
\item \textbf{Relative branch:} Each input is converted into PMT‐relative coordinates (one set per PMT), and---together with a polar/equatorial flag---is fed into $1\times1$ convolutional layers.\ This exploits the fact that all PMTs of the same type share identical responses.
 \item \textbf{Absolute branch:} The raw (absolute) inputs pass through dense layers that can learn symmetry-breaking effects, such as shadowing by the data cable.
\end{itemize}
\begin{figure}[p]
  \centering
  \begin{subfigure}{\linewidth}   
    \textbf{(a)}
    \centering
    \includegraphics[width=0.9\linewidth]{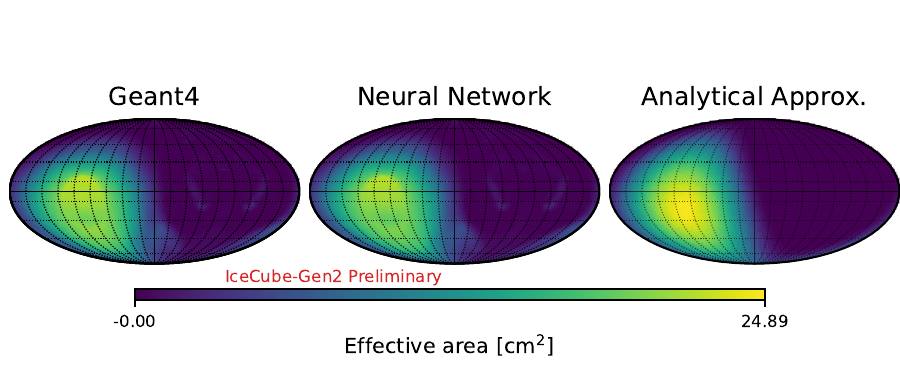}
  \end{subfigure}

  \begin{subfigure}{\linewidth}
    \textbf{(b)}
    \centering
    \includegraphics[width=0.9\linewidth]{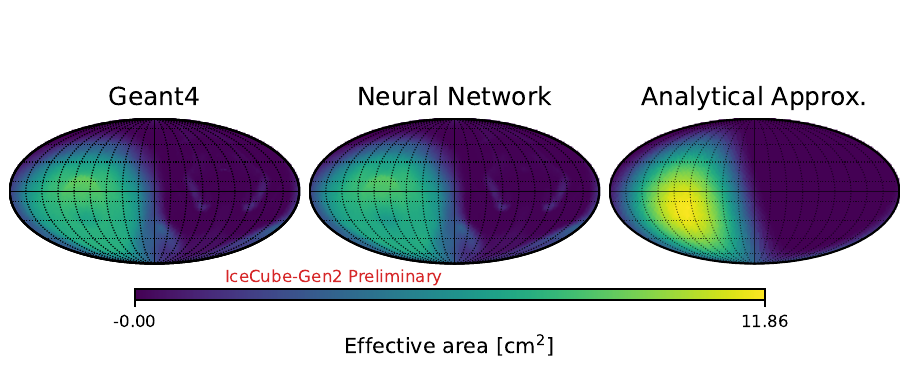}
  \end{subfigure}

  \caption{Omnidirectional maps of effective area $A_{\text{eff}}$ for a single equatorial PMT as a function of incident direction. From left to right: Geant4, Neural Network, and Analytical Approximation: \textbf{(a)} 400\;nm, \textbf{(b)} 550\;nm.}
  \label{fig:eff}
\end{figure}

\begin{figure}[p]
\centering 
\includegraphics[width=1.0\linewidth]{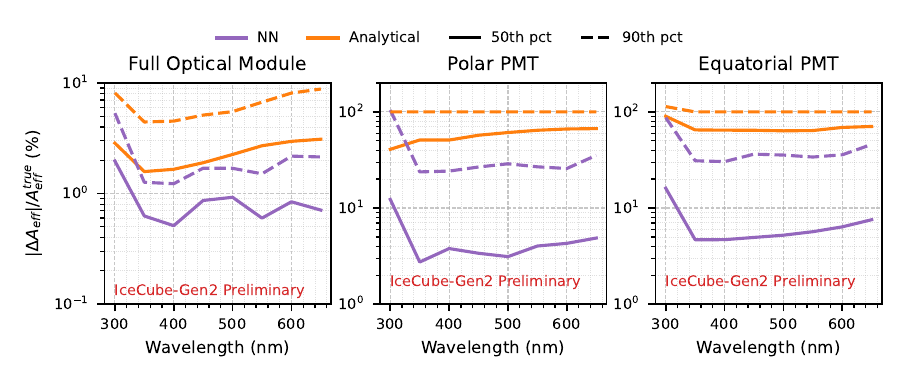}
\caption{50th (solid) and 90th (dashed) percentiles of the relative error
$\lvert A_{\mathrm{eff}}^{\text{Geant4}} - A_{\mathrm{eff}}^{\text{Approx}}\rvert
 / A_{\mathrm{eff}}^{\text{Geant4}}$ (in \%)
across 3072 isotropic directions, plotted versus wavelength.
“Approx” represents either the neural-network approach (purple) or the analytical approximation (orange).  Left to right: full optical module, polar PMT, and equatorial PMT.}
\label{fig:omnnperf}
\end{figure}

The output of the two branches is concatenated and a final logsoftmax activation function is used to have a normalized output that can be interpreted as a PDF.\ The training minimizes the Kullback–Leibler divergence \cite{Kullback1951-ye} between the network's PDF and the Geant4 reference.

This approach is not free from minor approximations: First, the ice within the sphere is modeled solely via its wavelength-dependent refractive index.\ Secondly, any photon that enters the sphere will never leave it.\ Finally, the time-of-flight between the sphere and detection is not considered; 99.8\% of photons were observed to be detected within 2\;ns, which is within the time uncertainty of the PMTs.
\subsection{Results}

The performance of the two methods, namely the neural-network approach and the analytical approximation, is evaluated using the effective area, $A_{\text{eff}}$.\ For a plane wave of photons with cross-sectional area $A_{\text{beam}}$, direction $(\theta,\phi)$, and wavelength $\lambda$, it is defined as:
\begin{equation}
    A_{\text{eff}}(\theta,\phi,\lambda)=\frac{N_{\text{det}}}{N_{\text{emit}}}\cdot A_{\text{beam}},
\end{equation}
where $N_{\text{emit}}$ is the amount of photons in the beam and $N_{\text{det}}$ the amount of detected photons.\ Figure~\ref{fig:eff} shows $A_{\text{eff}}$ versus incident direction for an equatorial PMT at 400\;nm (top) and 550\;nm (bottom), comparing Geant4, the neural‐network approach, and the analytical approximation.\ The neural network reproduces the asymmetries and wavelength dependence seen in the Geant4 maps.\ Figure~\ref{fig:omnnperf} further demonstrates that the NN yields a clear improvement in modeling angular acceptance---both for the full optical module and for the individual equatorial and polar PMTs---across all wavelengths.

In terms of inference speed, the network processes one million photons in 0.3 s on a GPU-roughly 200-300$\times$ faster than Geant4 on CPU---and should be acceptable for IceCube simulations.

\section{$\nu_{\mu}$ Charged-Current (CC) Directional Reconstruction Using Neural Networks}

One of the main goals of IceCube, and of IceCube-Gen2 in the future, is to localize the sources of high-energy neutrinos in the sky.\ To this end, $\nu_{\mu}$ CC events are particularly valuable, as they produce a track-like topology within the in-ice optical array that can be reconstructed with great angular resolution. Classical maximum-likelihood reconstructions \cite{Schatto2014-al} rely on high-dimensional lookup tables with ice-symmetry approximations due to computational constraints; however, those symmetries are broken by the new multidirectional optical modules.\ Neural networks natively handle this added complexity without approximation, making them ideal for next-generation DOM reconstructions.

\subsection{Implementation of Neural Network-based Reconstruction Techniques}

The following studies were carried out using the GraphNet framework \cite{Sogaard2023}.\ To this end, about 6.5 million $\nu_{\mu}$ CC events---which pass Monte Carlo truth-based quality cuts to remove poor quality events---were produced in the energy range from 1\;TeV to 50\;PeV.

High-energy events can produce millions of pulses---charge-time tuples reconstructed from PMT waveforms---which presents a computational challenge.\ We therefore followed the strategy in \cite{DNN_paper} and computed summary statistics for each PMT's pulse list.\ For each PMT, the inputs of the networks are its pulse summary statistics, its position, its orientation, and some flags such as module type (IceCube or IceCube-Gen2) and saturation status.

So far, we have found the GraphNet implementation of the IceCube Kaggle competition–winning solution \cite{bukhari2023icecubeneutrinosdeep} to be our best-performing model.\ It combines graph neural networks (GNNs) with a transformer architecture, although we replaced the latter with the Performer approach \cite{performer}, which uses an attention mechanism that scales approximately linearly with the input sequence length instead of quadratically. In fact, we observe a 4$\times$ speedup in training---a critical gain given that full model training can take several weeks on a GPU. 

Two approaches were followed for the directional reconstruction of $\nu_{\mu}$ CC:
\begin{itemize}
    \item \textbf{3D von Mises-Fisher (3D-vMF):} The point-spread function is assumed to follow the 3D-vMF distribution. Accordingly, the network outputs four values: the three-component reconstructed direction vector and the concentration parameter~\(\kappa\). These are then used to evaluate the 3D-vMF likelihood, and the network is trained by minimizing the negative log-likelihood of that distribution.
    \item \textbf{Conditional normalizing flows:} First introduced for IceCube in \cite{thorstenicrc}, they represent a new paradigm in neutrino‐telescope reconstruction. The goal is to predict the full posterior PDF of an event’s direction conditioned on the detector response. To this end, we feed the detector response into the model described above and then use an MLP to map its output onto the flow parameters. The network is trained by minimizing the negative log-likelihood of the predicted posterior PDF evaluated at the true direction. In our study, we used nine 2D-spherical flows from \cite{jammy_flows}.
\end{itemize}

\subsection{Results}

\begin{figure}[!t]
\centering 
\includegraphics[width=0.7\linewidth]{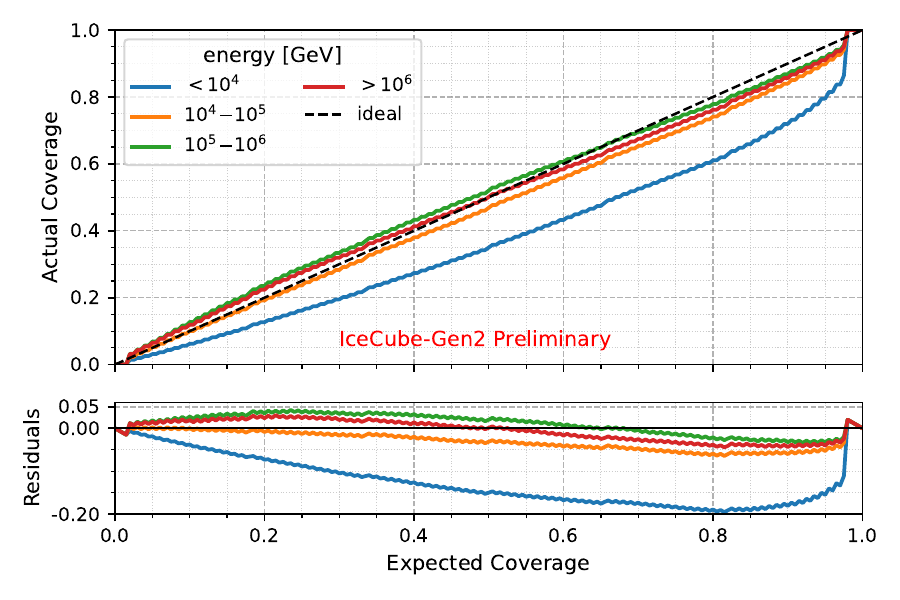}
\caption{Predicted vs. observed coverage for conditional normalizing flows for different deposited energies.}
\label{fig:cov}
\end{figure}

One of the most interesting aspects of conditional normalizing flows is their ability to predict asymmetric confidence contours.\ Figure~\ref{fig:cov} compares the nominal (expected) coverage of the predicted contours with their empirical coverage--—the fraction of events whose true direction lies inside the contour--—across different deposited-energy bins.\ The contours agree to within 5\;\% for energies above 10\;TeV but exhibit undercoverage at lower energies.\ This undercoverage indicates overfitting in the low-energy regime, which could be mitigated by augmenting the training set with additional low-energy events or by applying appropriate weighting during training.

\begin{figure}[!t]
\centering 
\includegraphics[width=0.9\linewidth]{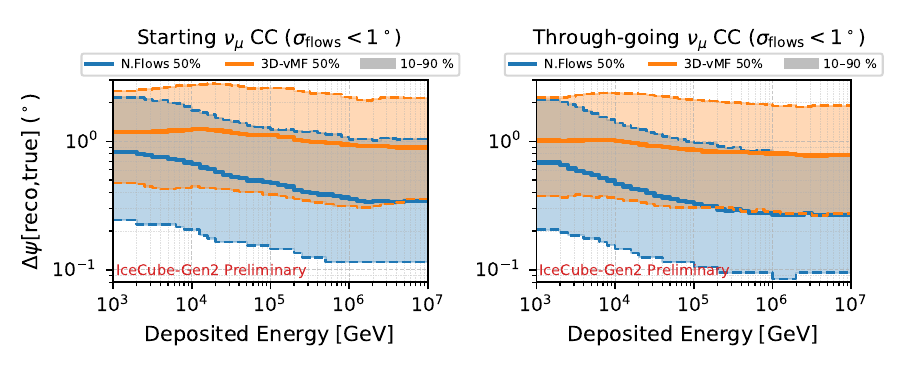}
\caption{Angular resolution as a function of deposited energy for the different approaches for starting and through-going $\nu_{\mu}$ CC.}
\label{fig:rec}
\end{figure}

A von Mises–Fisher approximation to the flows’ posterior PDF was performed as explained in \cite{thorstenicrc}, yielding a concentration parameter~\(\kappa\).\ For large~\(\kappa\), the 2-D Gaussian standard deviation can be approximated as \(\sigma = 1/\sqrt{\kappa}\).\ We observed that this uncertainty estimate agrees well with expectations for the normalizing flows approach, but not for the direct 3D-vMF method.\ A quality cut \(\sigma_{\mathrm{flows}}<1^\circ\) was applied in Figure~\ref{fig:rec}, which shows the angular resolution of \(\nu_{\mu}\) CC events starting inside the detector volume (left) and outside (right) as a function of deposited energy.\ This cut retains about 40\;\% (55\;\%) of starting events from trigger level at 10\;TeV (1\;PeV) and 60\;\% (65\;\%) of through-going events at the same energies.\ Conditional normalizing flows deliver the best overall performance, with especially promising resolutions for starting \(\nu_{\mu}\) CC.

\section{Noise Cleaning with Graph Neural Networks}
Radioactive decays in the glass of the optical modules generate background noise.\ The classical IceCube noise-cleaning algorithm filters reconstructed charge–time pulses according to their spatial and temporal distance from an initial set of seed pulses—--this is known as Seeded RT (SRT) cleaning.\ Although this technique performs well for IceCube DOMs, its efficiency is poorer for next-generation DOMs, which exhibit higher noise rates and several PMTs.

We followed a similar approach to that of \cite{eller2023sensitivityicecubeupgradeatmospheric} and trained a graph neural network to perform the cleaning.\ Our baseline is an adaptation of the DynEdge GNN model \cite{dynedge} implemented in GraphNet \cite{Sogaard2023}.\ Each event is treated as a point-cloud graph in which every pulse corresponds to a node whose features include the PMT position and orientation, as well as the pulse charge and time. The network is trained to binary-classify pulses as either physics or noise. 

\begin{figure}[h]
  \centering
  \begin{minipage}[b]{0.46\linewidth}
    \centering
    \includegraphics[width=\linewidth]{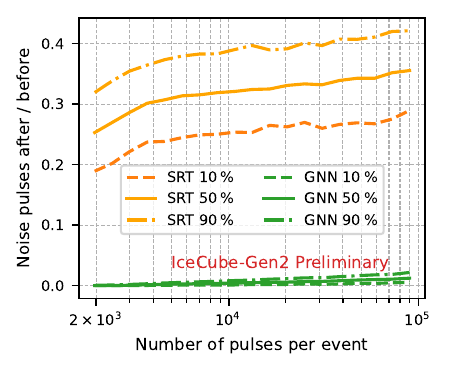}
    
    \vspace{0.3em}  
    \footnotesize\textbf{(a)}
    \label{fig:noise_noise}
  \end{minipage}
  \hfill
  \begin{minipage}[b]{0.46\linewidth}
    \centering
    \includegraphics[width=\linewidth]{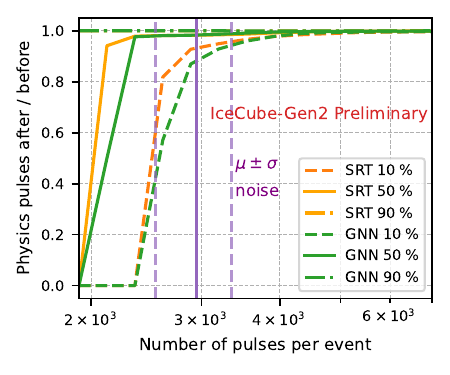}
    
    \vspace{0.3em}
    \footnotesize\textbf{(b)}
    \label{fig:noise_physics}
  \end{minipage}

  \caption{Remaining noise \textbf{(a)} and signal \textbf{(b)} pulse fractions versus total pulses per event, comparing the GNN-based cleaning to classical SRT; vertical purple lines in \textbf{(b)} mark the mean expected noise level $\pm$ 1 $\sigma$. }
  \label{fig:noise}
\end{figure}

The model was trained on about two million simulated $\nu_{\mu}$ CC and NC events, using only events that contained at most $10^{5}$ pulses due to computational resources.\ Figure~\ref{fig:noise} shows the performance of an independent sample of unseen $\nu_{\mu}$ CC and NC events.\ For benchmarking, the radial and temporal cuts of the standard IceCube SRT cleaning were scaled proportionally to the larger inter-string spacing expected for IceCube-Gen2.

The GNN-based approach suppresses more than 99\;\% of noise pulses for events containing up to \(10^5\) pulses, versus roughly 70\;\% for the classical SRT algorithm. Although SRT retains slightly more physics pulses in events with few pulses, the difference is small: at seven physics pulses per event, the median SRT retention exceeds the GNN by 0.8 pulses, and at 24 physics pulses per event, the median difference is less than 0.5 pulses.

Note that waveform simulation is not currently performed for IceCube-Gen2 PMTs, and the impact of mixed noise and physics hits within a single waveform will be studied in the future, although significant deviations are not expected.

\section{Conclusion and Outlook}\label{sec3}

Three neural-network techniques for the IceCube-Gen2 optical array have been presented. They fully exploit the increased dimensionality and complexity of the new optical modules and outperform the classical baselines evaluated here.\ We showed that NN-based optical-module simulation captures both the asymmetry and the wavelength dependence of the modules’ photon angular acceptance, with fast GPU inference. In addition, NN models reconstruct $\nu_{\mu}$ CC events with sub-degree angular resolution while providing reliable asymmetric uncertainty contours. Finally, a GNN-based noise-cleaning algorithm removes > 99\;\% of noise pulses from radioactive decays in the IceCube-Gen2 DOM glass, compared with about 70\;\% for the classical SRT algorithm, while retaining comparable signal information. Future work will extend these techniques to event classification and energy reconstruction, aiming for a complete IceCube-Gen2 event-selection chain for analysis studies, and will explore model-compression methods to accelerate CPU inference.

\bibliographystyle{ICRC}
\setlength{\bibsep}{0pt plus 0.3ex}
\bibliography{references}

%

\clearpage

\section*{Full Author List: IceCube-Gen2 Collaboration}

\scriptsize
\noindent
R. Abbasi$^{16}$,
M. Ackermann$^{76}$,
J. Adams$^{21}$,
S. K. Agarwalla$^{46,\: {\rm a}}$,
J. A. Aguilar$^{10}$,
M. Ahlers$^{25}$,
J.M. Alameddine$^{26}$,
S. Ali$^{39}$,
N. M. Amin$^{52}$,
K. Andeen$^{49}$
% ...
\\
\\
% ...
$^{10}$ Universit{\'e} Libre de Bruxelles, Science Faculty CP230, B-1050 Brussels, Belgium \\
% ...
$^{16}$ Department of Physics, Loyola University Chicago, Chicago, IL 60660, USA \\
% ...
$^{21}$ Dept. of Physics and Astronomy, University of Canterbury, Private Bag 4800, Christchurch, New Zealand \\
% ...
$^{25}$ Niels Bohr Institute, University of Copenhagen, DK-2100 Copenhagen, Denmark \\
$^{26}$ Dept. of Physics, TU Dortmund University, D-44221 Dortmund, Germany \\
% ...
$^{39}$ Dept. of Physics and Astronomy, University of Kansas, Lawrence, KS 66045, USA \\
% ...
$^{46}$ Dept. of Physics and Wisconsin IceCube Particle Astrophysics Center, University of Wisconsin{\textemdash}Madison, Madison, WI 53706, USA \\
% ...
$^{49}$ Department of Physics, Marquette University, Milwaukee, WI 53201, USA \\
% ...
$^{52}$ Bartol Research Institute and Dept. of Physics and Astronomy, University of Delaware, Newark, DE 19716, USA \\
% ...
$^{76}$ Deutsches Elektronen-Synchrotron DESY, Platanenallee 6, D-15738 Zeuthen, Germany \\
$^{\rm a}$ also at Institute of Physics, Sachivalaya Marg, Sainik School Post, Bhubaneswar 751005, India \\
% ...

\subsection*{Acknowledgments}

\noindent
The authors gratefully acknowledge the support from the following agencies and institutions:
USA {\textendash} U.S. National Science Foundation-Office of Polar Programs,
U.S. National Science Foundation-Physics Division,
U.S. National Science Foundation-EPSCoR,
U.S. National Science Foundation-Office of Advanced Cyberinfrastructure,
Wisconsin Alumni Research Foundation,
Center for High Throughput Computing (CHTC) at the University of Wisconsin{\textendash}Madison,
Open Science Grid (OSG),
Partnership to Advance Throughput Computing (PATh),
Advanced Cyberinfrastructure Coordination Ecosystem: Services {\&} Support (ACCESS),
Frontera and Ranch computing project at the Texas Advanced Computing Center,
U.S. Department of Energy-National Energy Research Scientific Computing Center,
Particle astrophysics research computing center at the University of Maryland,
Institute for Cyber-Enabled Research at Michigan State University,
Astroparticle physics computational facility at Marquette University,
NVIDIA Corporation,
and Google Cloud Platform;
Belgium {\textendash} Funds for Scientific Research (FRS-FNRS and FWO),
FWO Odysseus and Big Science programmes,
and Belgian Federal Science Policy Office (Belspo);
Germany {\textendash} Bundesministerium f{\"u}r Forschung, Technologie und Raumfahrt (BMFTR),
Deutsche Forschungsgemeinschaft (DFG),
Helmholtz Alliance for Astroparticle Physics (HAP),
Initiative and Networking Fund of the Helmholtz Association,
Deutsches Elektronen Synchrotron (DESY),
and High Performance Computing cluster of the RWTH Aachen;
Sweden {\textendash} Swedish Research Council,
Swedish Polar Research Secretariat,
Swedish National Infrastructure for Computing (SNIC),
and Knut and Alice Wallenberg Foundation;
European Union {\textendash} EGI Advanced Computing for research;
Australia {\textendash} Australian Research Council;
Canada {\textendash} Natural Sciences and Engineering Research Council of Canada,
Calcul Qu{\'e}bec, Compute Ontario, Canada Foundation for Innovation, WestGrid, and Digital Research Alliance of Canada;
Denmark {\textendash} Villum Fonden, Carlsberg Foundation, and European Commission;
New Zealand {\textendash} Marsden Fund;
Japan {\textendash} Japan Society for Promotion of Science (JSPS)
and Institute for Global Prominent Research (IGPR) of Chiba University;
Korea {\textendash} National Research Foundation of Korea (NRF);
Switzerland {\textendash} Swiss National Science Foundation (SNSF).

\end{document}